\documentclass[aip,rsi,reprint]{revtex4-1}

\usepackage{graphicx}
\usepackage{bm}

\begin{document}

\addtolength{\textheight}{1cm}

\newcommand{\Mx}{\ensuremath{{M}_{{x}}}}

\newcommand{\Hx}{\ensuremath{{H}_{{x}}}}

\newcommand{\Hp}{\ensuremath{H_{{p}}}}
\newcommand{\Hb}{\ensuremath{{H}_{{b}}}}
\newcommand{\Ht}{\ensuremath{{H}_{{t}}}}
\newcommand{\Heff}{\ensuremath{{H}_{\text{eff}}}}

\newcommand{\unit}[2]{$#1\,\mathrm{#2}$}

\newcommand{\Py}{$\text{Fe}_{20}\text{Ni}_{80}$ }

\title{Time resolved X-ray Resonant Magnetic Scattering in reflection geometry} 



\author{Stefan Buschhorn}
\email[corresponding author: ]{stefan.buschhorn@ruhr-uni-bochum.de}

\author{Frank Br\"ussing}
\author{Radu Abrudan}
\author{Hartmut Zabel}

\affiliation{Institut f\"ur Experimentalphysik/Festk\"orperphysik, Ruhr-Universit\"at Bochum, Germany}


\date{\today}

\begin{abstract}
We present a new setup to measure element-selective magnetization dynamics using the ALICE chamber (RSI \textbf{74}, 4048 (2003)) at the BESSY II synchrotron at the Helmholtz-Zentrum Berlin. A magnetic field pulse serves as excitation, and the magnetization precession is probed by element selective X-ray Resonant Magnetic Scattering (XRMS). With the use of single bunch generated x-rays a temporal resolution well below \unit{100}{ps} is reached. The setup is realized in reflection geometry and enables investigations of thin films described here, multilayers, and laterally structured samples. The combination of the time resolved setup with a cryostat in the ALICE chamber will allow to conduct temperature-dependent studies of precessional magnetization dynamics and of damping constants over a large temperature range and for a large variety of systems in reflection geometry. 
\end{abstract}


\maketitle 

\section{Introduction}
\label{sec:Intro}
\enlargethispage{5mm}

Nanomagnetism on short timescales has attracted much interest in recent years for both a fundamental understanding and for technological reasons. The time scales reach from seconds for fluctuations of magnetization in nanoparticles and ultrathin films \cite{wernsdorfer_PRL_97, shpyrko_N_07}, micro- to nanoseconds for magnetization reversal processes via domain wall motion \cite{klaui_JoPCM_08}, picoseconds for precessional dynamics \cite{kalarickal_JoAP_06}, to femtoseconds for demagnetization processes via heat pulses \cite{beaurepaire_PRL_96, stamm_nm_07, durr_MITo_09}. The precessional dynamics that occurs in response to a change of the external magnetic field direction is of particular interest, as it constitutes the basic step to a complete magnetization reversal. The damped precessional motion about the new field direction is entirely governed by the Landau-Lifshitz-Gilbert equation (LLG).  The precessional dynamics can be studied either by driving the ferromagnetic system into resonance via microwave excitation (FMR) \cite{farle_RoPiP_98, heinrich_STiMP_08}, where the width of the resonance line is a measure of the damping constant, or in real time via a step or impulse excitation, where the system subsequently relaxes into the effective field direction with a characteristic time constant \cite{nibarger_APL_03, silva_JoAP_99}. While FMR experiments on nanomagnetic systems are well established \cite{bland_book_05}, pulse field excitation experiments are less common. Pioneering work was done by Silva and Gerrits using an Auston switch for generating a strong current pulse, which, in turn, produces a magnetic field pulse within the sample \cite{gerrits_n_02, gerrits_RoSI_06}. Using time resolved Magneto-Optical Kerr Effect (tr-MOKE) \cite{bauer_APL_00, neudert_PRB_05, hubert_book_98} the precessional switching can be resolved in real time. For magnetic alloys and for magnetic heterostructures it is desirable to have a method which not only has sufficient time resolution but also provides element selective information. For domain wall and vortex dynamics, time-resolved Photo\-emission Electron Microscopy (tr-PEEM) \cite{kaiser_JoPCM_09, vogel_APL_03} and time resolved Scanning Transmission X-ray Microscopy (tr-STXM) \cite{bocklage_PRB_08, vanwaeyenberge_n_06} experiments are powerful element selective tools using resonant absorption. These techniques, however, image domains rather than individual moments. Time resolved X-ray Resonant Magnetic Scattering  (tr-XRMS) is the method of choice for the investigation of element specific precessional motion and has been demonstrated for the first time by Bailey \emph{et al.} \cite{bailey_PRB_04} in the time domain and in reflection geometry. These experiments where later accompanied by transition geometry experiments with FMR \cite{arena_RoSI_09, martin_JoAP_09} and pulsed excitation \cite{martin_JoAP_08}.

Here we describe a newly developed tr-XRMS setup that allows excitation of nanomagnetic systems using field pulses on the 100 picosecond timescale.  The free precessional response as well as the precessional damping over several nanoseconds is then followed for each element by tuning the photon energy to the different x-ray resonance absorption edges. This real time method also enables to address the low frequency limit of precessional motion that is still challenging in the frequency domain. In the following we first describe the experimental setup of our system and then provide results from a \Py (Py) thin film sample.

\section{Experimental Setup}
\label{sec:Setup}
\begin{figure}[b]
\includegraphics[width=\columnwidth]{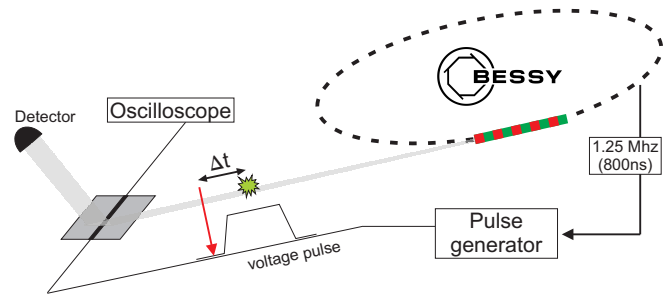}
\caption{A schematic of the pump-probe setup. The delay between voltage pulse and photon pulse is varied, and the reflected intensity is measured as a function of delay time. }
\label{fig:RSI-principle-scheme}
\end{figure}

In our setup we combine the element selective technique of XRMS with the pulsed structure of the x-rays provided by a synchrotron storage ring in order to enable time-resolved experiments, as depicted in Figure~\ref{fig:RSI-principle-scheme}.

The sample is excited by a magnetic field pulse, and after a fixed delay the instant magnetization is probed by the x-ray pulse. By controlling the delay between pump and probe in steps as small as \unit{10}{ps}, we monitor the instant magnetization at fixed times after the onset of the excitation. The resulting delay scan yields an element and time resolved sequence of the precessional dynamics in the sample. We emphasize that tr-XRMS measures the individual magnetic moments in contrast to tr-MOKE \cite{neudert_PRB_05}, tr-PEEM \cite{kaiser_JoPCM_09, vogel_APL_03} or tr-STXM\cite{bocklage_PRB_08, vanwaeyenberge_n_06} experiments, where typically the dynamics of domain structures like Landau patterns or vortex core reversal is monitored. Due to the photon-in photon-out technique, there is no charging effect to worry about as may be the case for PEEM experiments.

The experiments were carried out at the UE52 beamline at the BESSY II synchrotron at the Helmholtz-Zentrum Berlin (HZB), using the ALICE diffractometer\cite{grabis_RoSI_03} as end station. The undulator beamline provides x-rays of variable energy and polarization, including circularly polarized light at the resonant energies of the 3d transition elements. During the single bunch operation mode of the synchrotron, the intensity of the undulator is still sufficiently high (about one order of magnitude less as compared to multibunch mode) to render time resolved experiments feasible. The ALICE chamber is a versatile two-circle diffractometer with horizontal scattering geometry and a broad field and temperature range is accessible at the sample position. Using circularly polarized light in reflection geometry at the element-specific resonant energies, we probe the horizontal - or \Mx~(collinear with the beam) component of the magnetization (equivalent to the longitudinal MOKE geometry). In this way element-specific information on the static magnetization profile of various magnetic heterostructures is accessible. 

\begin{figure}
\includegraphics[width=0.9\columnwidth]{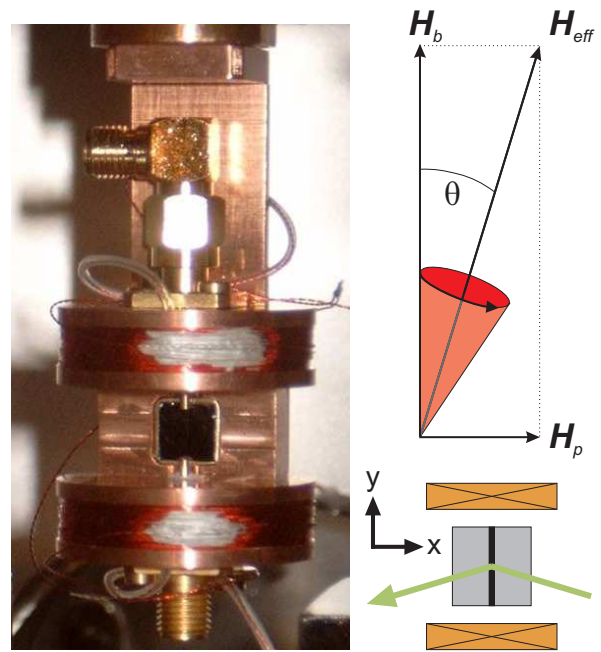}
\caption{Picture of the sample holder with coils and hf connectors. The black rectangle is a Si substrate, positioned from the backside. An M8 screw at the top connects the sample holder with the cryostat (upmost part). The right panel is a schematic of the stripline sample geometry with the bias field \Hb~parallel to the stripline and the pulsed field \Hp~provided by a current through the stripline perpendicular to it. The vector sum of both fields defines the effective field direction \Heff, about which the precessional magnetization dynamics occurs before it is aligned parallel to the \Heff~axis. In the time resolved resonant magnetic x-ray scattering experiment the x-component of the precessing magnetization \Mx~is detected as a function of time after pulse excitation. }
\label{fig:RSI-field-directions}
\end{figure}

For the time-resolved experiments we designed a special sample holder for the ALICE chamber serving the needs for an additional field in the direction perpendicular to the scattering plane and for the high frequency (hf) supply. A photograph of the sample holder together with a schematic outline of the setup is shown in Fig~\ref{fig:RSI-field-directions}.
This sample holder combines in a convenient way the capabilities of the diffractometer in terms of field, angular and temperature range with the possibility to conduct time resolved experiments at the very same sample. In order to ensure good thermal conductivity and to minimize temperature drifts the sample holder is fabricated from Cu. It includes a pair of coils to generate a bias field \Hb~perpendicular to the scattering plane, and SMA connectors to contact the sample to the hf wiring. Two pins serve as electrical contacts by just pressing a sample against them from the backside. These solder-free contacts enable simple and fast sample change from the backside of the sample holder without the necessity of disconnecting any hf wiring. The sample itself is glued on a small sledge and fixed through an opening from the backside of the sample holder. Due to its position it is always placed at the center of rotation once the sample holder is aligned, resulting in very short alignment times. A delta electronika constant current source delivers up to \unit{460}{mA} for the coils, resulting in a maximum bias field of $\approx 100\,\text{Oe}$. However, this field is only used to saturate the sample along the stripline prior to data aquision, as the heat dissipation of the coils generates a temperature drift and limits the permanent currents to about \unit{250}{mA}. The sample is a $7 \times 7\,\text{mm}^{2}$ Si substrate with a centered conducting  stripline of width \unit{50}{\mu m} and length \unit{7}{mm} as main conducting layer and a magnetic layer deposited on top.

\begin{figure}
\includegraphics[width=\columnwidth]{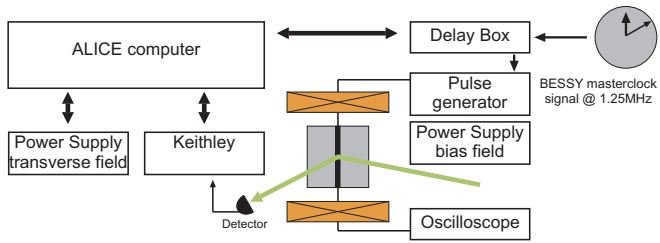}
\caption{Schematic outline of the electronics for time resolved experiments. The left part shows the standard ALICE equipment, the right part the electronics setup for the time resolved experiments. The double arrows mark GPIB communication.}
\label{fig:RSI-electronic-circuit}
\end{figure}

The main parts of the sample holder are sketched in the lower right panel of Figure~\ref{fig:RSI-field-directions}. Since we use circularly polarized light we are sensitive to the magnetization parallel to the \textit{x}-direction, thus the magnetic contribution to the reflected signal is $\text{I}_{\text{m}} \propto \bm{\sigma} * \textbf{M}$. The rotatable electromagnet in the ALICE chamber provides a magnetic field \Hx~parallel to the sample surface and in the scattering plane in order to measure element selective hysteresis loops 
for static characterization. A constant current applied through the stripline will then result in a shift of the hysteresis, as the magnetic field generated by the current will add to the external field \Hx~supplied by the electromagnet. The bias field \Hb, on the other hand, induces an easy-axis behavior parallel to the stripline and thus a hard-axis-behavior along the \textit{x}-direction. The fields present at the sample position are shown in the top right panel of Fig.~\ref{fig:RSI-field-directions}. Temporal resolution is obtained with a pump-probe technique during single bunch operation mode at BESSY II: Only one bunch travels in the storage ring, the photons generated arrive in pulses of \unit{50}{ps} width and a separation of \unit{800}{ns}, i.e. a repetition rate of \unit{1.25}{MHz}. The synchrotron masterclock provides a trigger signal at this frequency with a fixed yet arbitrary phase to the photons hitting the sample.  The photon pulse length sets the upper limit of the time resolution. Any processes that are faster than \unit{50}{ps} are not accessible with our setup. For the excitation of the precession the trigger signal from the BESSY masterclock is fed into an HP8130A pulse generator, which serves as a delay station and is controlled via a GPIB interface. The output signal is then connected to an AvTech pulse generator, generating a voltage step of \unit{10}{ns} length and a variable amplitude of up to \unit{10}{V} with a rise time of \unit{225}{ps} (10/90). This pulse is delivered to the sample via $50\,\Omega$ coaxial cables and finally fed into a \unit{20}{GHz} HP 54110D  oscilloscope during the measurements. All cables and electrical feedthroughs used in this setup are chosen for small damping loss in the frequency range up to \unit{18}{GHz} to sustain the high frequency components of the edges of the voltage pulse and to maintain a sharp rise and fall time of the field pulse. This is essential for observing free precessional motion in the sample instead of an adiabatic change of the magnetization vector in response to a change of the field direction. The voltage pulse directly results in a magnetic field pulse at the sample position, the shape of the voltage pulse observed at the scope is a direct measure for the shape of the magnetic field pulse. A schematics of the electronic circuit is shown in Fig.~\ref{fig:RSI-electronic-circuit}. 

The reflected intensity is detected by a GaAs photodiode together with a Keithley pico ammeter, as a function of delay time. Delay control and data acquisition is done via the SPEC software \footnote{Certified Scientific Software, Cambridge, MA 02439. http://www.certif.com}. The delay between the masterclock signal and the voltage pulse at the sample is controlled with the HP8310A via a GPIB interface. The minimum step size possible is \unit{10}{ps}, which is well below the intrinsic length of the photon pulse (\unit{50}{ps}).

For each delay step, the signal is integrated over for about \unit{30}{s}, and a reference signal is measured without applying a current pulse. The latter is important to monitor the background signal and to avoid stability effects due to the small size of the sample. As a whole scan takes about half an hour, stability of the sample with respect to the beam is an important issue. For striplines of only \unit{50}{\mu m} width we observed a very noisy signal that we assign to small variations of the sample position with respect to the beam. This resulted in intensity variations of up to 5\,\%, which is in the range of the signal to be measured and prevents any time resolved scans to be taken. Therefore, the width of the stripline should be on the same order as the beam size.

\section{Results and Discussion}
\label{sec:Results}

\begin{figure}[b]
\includegraphics[width=\columnwidth]{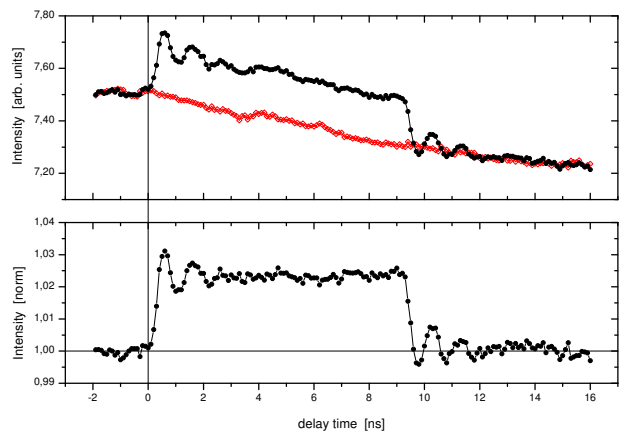}
\caption{The upper part shows the intensity detected as a function of delay time with (full) and without (open symbols) current pulse. Both intensities drop because of the ring decay. The lower graph shows the same delay scan where the intensity with pulsed excitation is normalized to the one without. The oscillations at the leading and trailing edges are clearly recognized.}
\label{fig:RSI-pulse-ref}
\end{figure}

\begin{figure}
\includegraphics[width=\columnwidth]{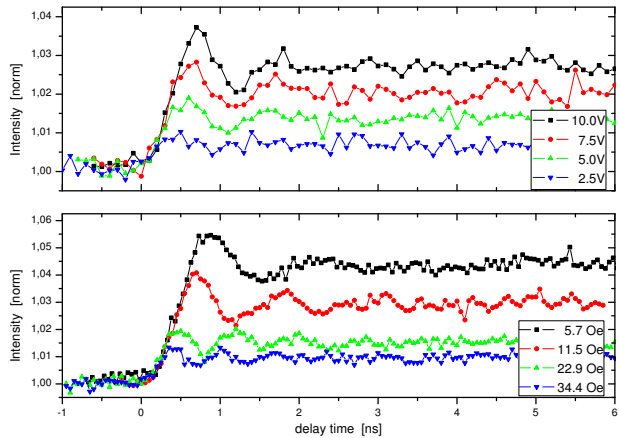}
\caption{The upper panel shows delay scans at four different amplitudes of the pulsed field at a constant bias field of \unit{11.5}{Oe}. The lower panel shows the effect of an increasing bias field at a constant pulse field. The change in step height is not linear with the bias field, because there is always a constant $H_{\mathrm{eff},x} = \Hp$ present.}
\label{fig:RSI-field-and-amplitude-variation}
\end{figure}

As a test sample we present results of a polycrystalline \unit{25}{nm} thick Py layer on top of a \unit{50}{nm} thick Cr stripline. The structure was deposited on a Si substrate through a mask resulting in a \unit{600}{\mu m} wide stripline. All scans shown in Fig.~\ref{fig:RSI-pulse-ref} and \ref{fig:RSI-field-and-amplitude-variation} were taken at the Fe $L_{3}$ resonant edge at an incident and exit glancing angle of $7\,^{\circ}$. Prior to each delay scan the sample is saturated along the stripline and subsequently the bias field is released to a fixed value. The magnetization is then considered to be aligned along the \textit{y}-direction, yielding no magnetic asymmetry in the reflected intensity. The change in reflected intensity is then monitored as a function of delay time between excitation and probe. Figure~\ref{fig:RSI-pulse-ref} shows a typical result for a bias field of \unit{11.5}{Oe} and a voltage pulse of \unit{10}{V} amplitude and  \unit{10}{ns} length. 
The overall step in the intensity arises from the magnetic field pulse, and the step height represents the new equilibrium direction for the magnetization being aligned along  \Heff. In a first approximation this is given by the vector sum of \Hb~and \Hp, neglecting any other contributions (compare the right part of Fig~\ref{fig:RSI-field-directions}). In this picture the angle $\theta$ between \Heff~ and \Hb~ can be determined from the intensity change at the step. The change of the direction of the magnetic field with the rising edge of the current pulse leads to damped free precession about \Heff. In the reflected intensity we observe the projection of the magnetization precession into the \textit{x}-direction, which appears as a damped harmonic oscillation. The damped oscillations at both leading and trailing edge of the excitation are clearly visible, lasting for a few nanoseconds. Anisotropies are neglected in this picture but may become important for a more detailed analysis of the data. To proof that the observed oscillations are not by any means a geometric effect, we have changed photon helicity, bias field direction and current pulse direction, obtaining comparable results in all cases. 

Delay scans at a fixed bias field of \unit{11.5}{Oe} for four different pulse amplitudes are shown in the top panel of Fig. 5 at the rising edge of the field pulse. The overall step height decreases with decreasing pulse amplitude as  expected since the angle $\theta$ between \Heff~ and \Hb~ and the projection of $\mathbf{M}$ into the \textit{x}-direction becomes smaller for smaller \Hp. This almost linear reduction leads to the assumption that the pulsed field is small compared to the bias field. The lower panel of Fig.~\ref{fig:RSI-field-and-amplitude-variation} shows the effect of different bias fields at fixed pulse amplitude. Again, the step height decreases with increasing bias fields as $\theta$ decreases with increasing bias field. The data also clearly show an increasing oscillation frequency, which is expected for increasing effective field according to the Kittel formula \cite{kittel_book_96}. Finally, in both data sets we notice that the initial slope upon the excitation is similar for all parameters chosen, although the final step height is different. The time where \Mx~first transits to the new equilibrium increases with the step height.  The observed frequencies are in the low GHz regime, which is reasonable for Py. We have obtained equivalent results for the Ni moments in Py, when tuning the x-ray energy to the Ni $L_{3}$ edge. This proves that, with the setup described here, we are able to measure element-resolved precessional magnetization dynamics; a detailed analysis of our data will be reported elsewhere.  

\section{Conclusion and Outlook}
\label{sec:Conclusion} 	 	
We have constructed an add-on sample holder for the ALICE chamber, which together with the electronics enables time and element resolved experiments in reflection geometry with a time resolution of less than 100\,ps. A pulsed magnetic field triggers a magnetization precession $\mathbf{M}$(t) about a new effective field direction \Heff, the projection of which into the scattering plane \Mx(t) is detected as a damped oscillation after defined time delays. This stroboscopic detection mode allows time scans from \unit{100}{ps} up to a few ns. The expected dependencies of the \Mx~component on both the pulsed magnetic field amplitude and the bias field strength are clearly recognized for a Py thin film sample. Furthermore, we observe the expected frequency increase for increasing bias field. Future experiments will focus on the magnetization dynamics and damping in multilayer structures and laterally patterned samples. The reflection geometry is ideally suited for these kinds of samples, as the substrate can be chosen arbitrarily, and furthermore depth information is obtained via variation of the scattering vector. 

\begin{acknowledgments}
We would like to thank Dario Arena and Alexei Nefedov for helpful discussions. This work was supported by the BMBF Verbundforschung (05KS7PC1), which is gratefully acknowledged. We are also thankful to HZB Berlin supported by BMBF for travel funds under 05 ES3XBA/5.  
\end{acknowledgments}

\bibliography{RSI_literature}

\end{document}